\documentclass[aps,manuscript,aps, prl, reprint, showpacs]{revtex4-1}
\usepackage{lmodern}

\usepackage[T1]{fontenc}
\usepackage{color}
\usepackage{amssymb}
\usepackage{graphicx}
\usepackage{esint}
\usepackage[unicode=true,pdfusetitle,
 bookmarks=true,bookmarksnumbered=false,bookmarksopen=false,
 breaklinks=false,pdfborder={0 0 1},backref=false,colorlinks=true]
 {hyperref}

\makeatletter
 
 \@ifundefined{textcolor}{}
 {%
   \definecolor{BLACK}{gray}{0}
   \definecolor{WHITE}{gray}{1}
   \definecolor{RED}{rgb}{1,0,0}
   \definecolor{GREEN}{rgb}{0,1,0}
   \definecolor{BLUE}{rgb}{0,0,1}
   \definecolor{CYAN}{cmyk}{1,0,0,0}
   \definecolor{MAGENTA}{cmyk}{0,1,0,0}
   \definecolor{YELLOW}{cmyk}{0,0,1,0}
 }

\makeatother

\begin{document}

\title{Exciton Dynamics in Carbon Nanotubes: From the Luttinger Liquid to
Harmonic Oscillators}

\author{M. C. Sweeney and J. D. Eaves}

\affiliation{Department of Chemistry and Biochemistry, University of Colorado,
Boulder, Colorado 80309, USA}
\begin{abstract}
We show that the absorption spectrum in semiconducting nanotubes can
be determined using the bosonization technique combined with mean-field
theory and a harmonic approximation. Our results indicate that a multiple
band semiconducting nanotube reduces to a system of weakly coupled
harmonic oscillators. Additionally, the quasiparticle nature of the
electron and hole that comprise an optical exciton emerges naturally
from the bosonized model. 
\end{abstract}

\pacs{78.67.Ch, 71.10.Pm}

\maketitle
\begin{figure}[h]
\includegraphics[scale=0.57]{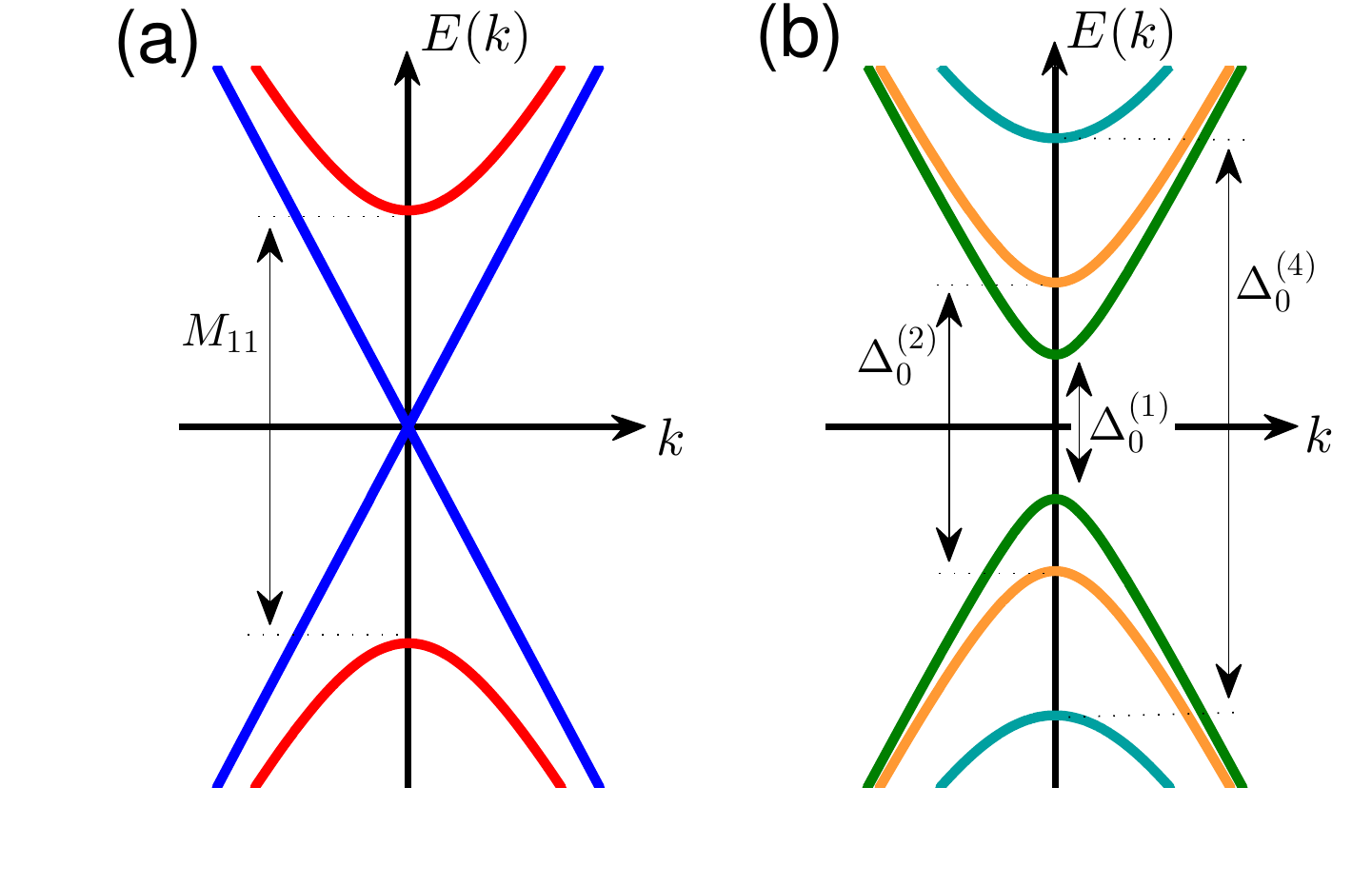}

\caption{\label{fig:Dispersion}(Color online) Dispersion relationships from
the zone-folding approximation for a metallic tube, (a), and for a
semiconducting tube, (b), in the vicinity of the low-energy excitations.
$M_{11}$ indicates the position of the van Hove singularities in
the metallic tube, $\Delta_{0}^{(n)}$ indicate the bare gaps in the
semiconducting tube. Note that within the context of our theory the
bare gap is phenomenological, and may indeed be present in SWNTs that
would be metallic in the zone-folding approximation.}
\end{figure}

Many of the properties of single-walled carbon nanotubes (SWNTs) are
deeply rooted in the physics of strongly interacting electrons in
low spatial dimensions \cite{ando_excitons_1997}. In SWNTs, the low-energy
fluctuations in the electron density are dominated by one-dimensional
excitations of the electrons in the $\pi$-energy bands. SWNTs can
transport electrons like a nearly ideal one-dimensional conductor,
but more like molecules than solid-state materials, display sharp
lines in their absorption spectra \cite{Brus2010,raey}. These two
faces: part solid-state and part molecular, make SWNTs unique nanoscale
systems. The bands made from these orbitals are characterized by a
wavevector and band index (Fig.~\ref{fig:Dispersion}). The lowest
energy bands can be semiconducting or metallic, depending on the chirality
of the tube \cite{saito1998physical}. Because the susceptibility
is related to the density-density correlation function through the
continuity equation, optical excitations probe the quantum mechanical
electronic density fluctuations of the SWNT. A consistent and comprehensive
picture of the optical excitations and electronic dynamics in semiconducting
SWNTs is important at a fundamental level. Such a picture may have
practical consequences in certain applications, because one might
exploit novel properties that emerge from strong electron-electron
interactions. Because semiconducting SWNTs absorb strongly in the
near-IR of the spectrum, they are also promising candidates for solar
energy applications. In such applications one needs to understand
not just how and where the SWNT will absorb light, but about the subsequent
\emph{electronic dynamics} following absorption, such as interband
scattering, Auger recombination, and multiple exciton generation (MEG)
\cite{perfetto_theory_2008,Schneck2011,Graham2011}.

The quasiparticle approach has made some remarkably accurate predictions
for the absorption spectra of SWNTs \cite{Spataru2003,Deslippe2007}.
In this picture, the absorption of a photon produces a Coulomb-bound
electron-hole quasiparticle state called an exciton \cite{Hybertsen1986,ando_excitons_1997,Spataru2005,Dresselhaus2007},
and the Bethe-Salpeter equation describes the electron-hole interaction.
As a consequence of the one-dimensional quantum confinement excitons
dominate the optical spectrum \cite{ando_excitons_1997,Wang2005}.
But transport phenomena in metallic SWNTs are predicted to be dominated
by fluctuations that have no intrinsic correlation length and lie
outside this quasiparticle paradigm \cite{KaneCoulomb1997,Balents2000a,raey}.
These fluctuations are described naturally within Luttinger liquid
(LL) theory \cite{Giamarchi2004,Voit1995}. Here too, some of the
predicted transport properties have been reported in experiments \cite{bockrathLuttingerLiquid1999,Kim2007}.
The LL approach has also been applied to study transport in semiconducting
SWNTs \cite{Levitov2003}, but only recently been applied to study
optical excitations in SWNTs \cite{konik2011}. 

In this letter we analyze optical excitations and relaxation dynamics
in SWNTs by applying Luttinger liquid theory to SWNTs with multiple
bands. We present a mean-field and variational treatment of a bosonized
multiband gapped Luttinger liquid model that clarifies some outstanding
issues about the nature and dynamics of electronic excitations in
a gapped one-dimensional system. The resulting theory makes predictions
for optical transition energies, the $E_{ii}$, in SWNTs that agree
well with experimental measurements, and we comment on the relaxation
dynamics of the excitations between bands. We find that the Luttinger
liquid is fragile to gaps of \emph{any} size, and that once a gap
is introduced into the SWNT, electron-electron interactions widen
it and the correlation length for the electronic density fluctuations
becomes finite. At any finite gap these fluctuations resemble particles
(electrons), antiparticles (holes), and an excitonic state that can
all be classified according to their topological charge. While terms
that give rise to MEG and Auger relaxation are in the model, these
are interband processes. Such processes become progressively weaker
relative to the intraband Coulomb interaction as the length to diameter
ratio for the SWNT gets larger, consistent with the findings in Ref.~\onlinecite{baer_can_2010}. 

\begin{figure}
\includegraphics[scale=0.58]{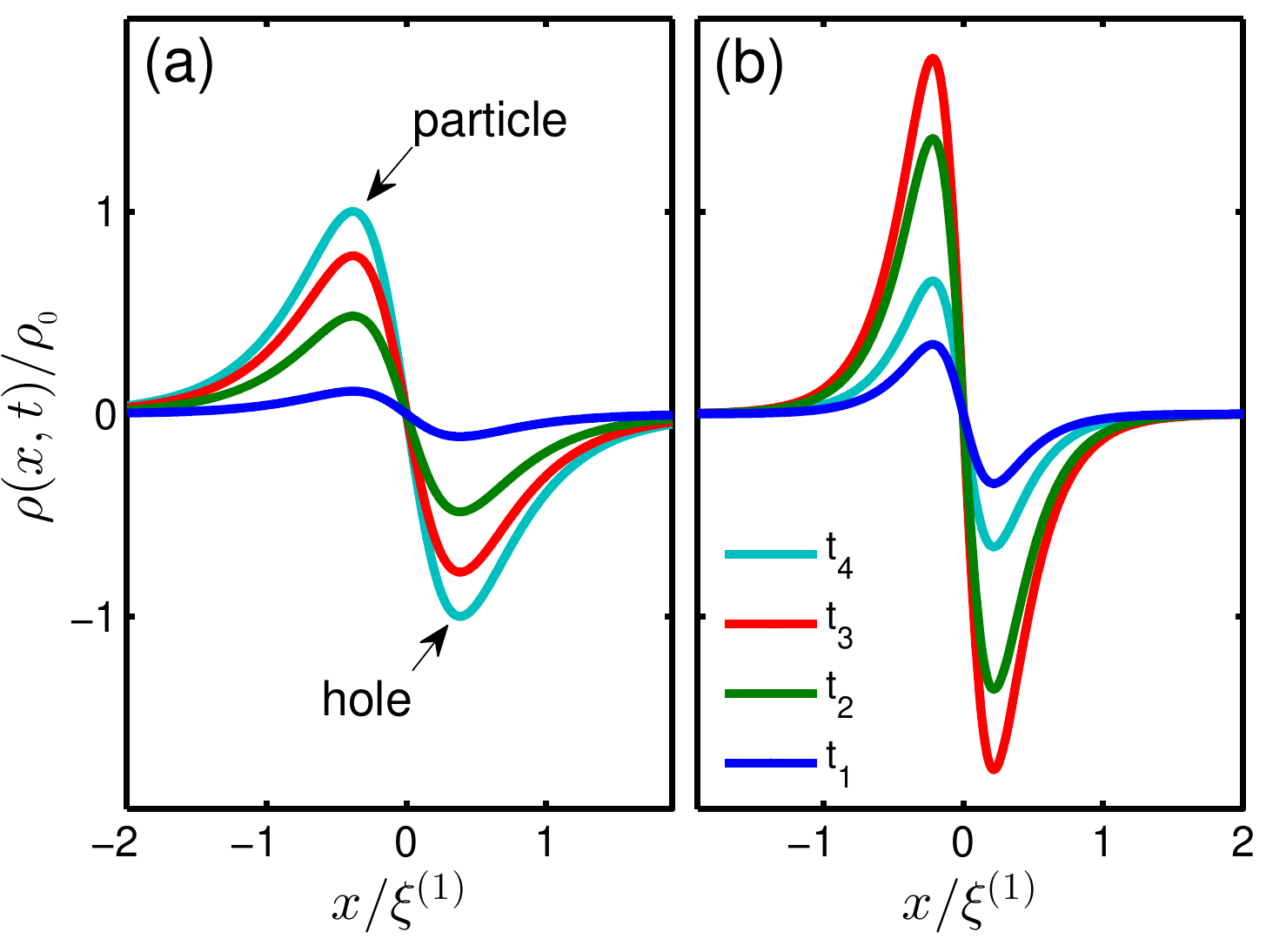}

\caption{\label{fig:ClassicalBreatherSolutions}(Color online) Classical breather
solutions to the scalar sine-Gordon model \cite{Rajaraman1996}, illustrating
the emergence of quasiparticle behaviors and the relative difference
between breathers/excitons for different bands \cite{SupplementaryMaterial}.
The electronic density is related to the total charge sector boson
field, $\rho\left(x,t\right)\propto\partial_{x}\theta_{c+}\left(x,t\right)$.
(a) The $n=1$ breather over the first quarter period of its ``orbit''.
The particle and hole that comprise the breather oscillate about the
zero position. (b) The $n=2$ breather at the same time intervals,
its period is roughly half of that for the first band, $T_{2}=0.57\, T_{1}$.
The solutions in this figure have $\theta_{c+}$ values centered about
a cosine minima, see Eq.~(\ref{eq:sG_Coupling}). The units are defined
by the $n=1$ peak density and particle/hole separation distance.}
\end{figure}

\begin{figure}
\includegraphics[scale=0.6]{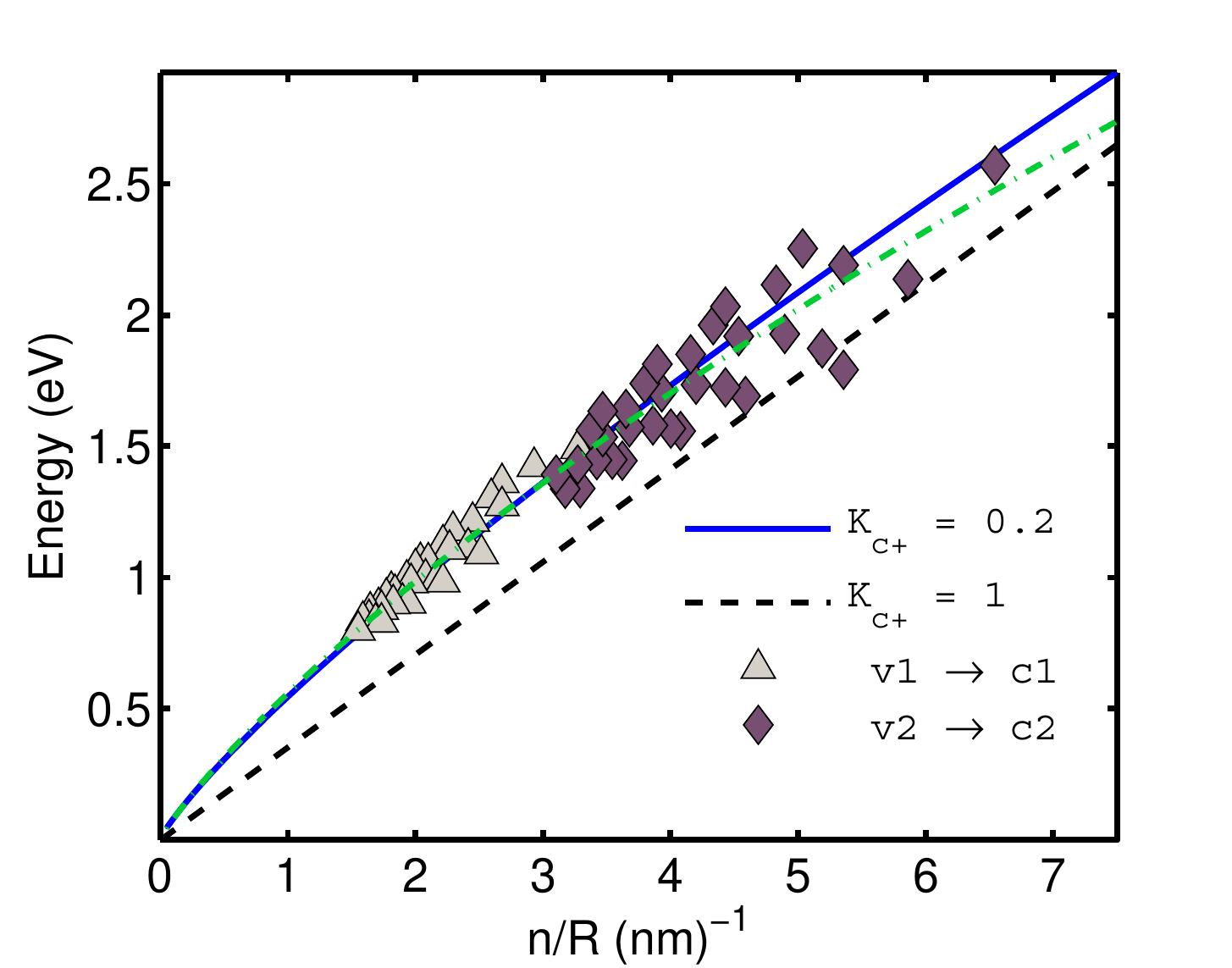}\caption{\label{fig:Comparison-to-Experiment}(Color online) Optical transition
energies for the first two bands, $E_{11}$ and $E_{22}$, for semiconducting
nanotubes. The solid blue line is from the harmonic approximation,
Eq.~(\ref{eq:HarmonicMassTerm}). The triangle and diamond data points
are experimental results from Ref.~\onlinecite{Bachilo2002}. The
green dot-dash line is from Ref.~\onlinecite{KaneElectron2004}.}
\end{figure}

The usual methods to solve for optical excitations in nanotubes begin
with dressed electron and hole states, typically at the level of Hartree-Fock
mean-field theory \cite{Dresselhaus2007,ando_excitons_1997}. In essence
this method takes the dressed states, or quasiparicles, of the gapped
bands as a reference Hamiltonian. The complete Hamiltonian with the
Coulomb interaction is solved approximately with respect to the reference
Hamiltonian. The method we employ shares some similarities to this
approach. For the gapless case, bosonizing the free-particle energy
with the (forward scattering) Coulomb interaction gives an exactly
solvable Hamiltonian. The solution is a Luttinger liquid, a system
that lies along a line of quantum critical points, where density-density
correlation functions follow a nonuniversal power law. In this work
the LL Hamiltonian is used as the reference. Relative to this reference
an intrinsic gap is relevant in the renormalization group (RG) sense.
Interband scattering is marginal. At long wavelengths, fluctuations
widen a bare gap, while the interband scattering matrix element tends
towards a constant value \cite{Lin1993}. The gap term is the strongly
interacting part of the Hamiltonian in the bosonized representation
and cannot be treated perturbatively \cite{Levitov2003,konik2011}.
We solve for the intraband gapped LL using mean-field theory and a
variational harmonic Ansatz. The excitonic nature of the excitations
emerges from the mean-field result (Fig.~\ref{fig:ClassicalBreatherSolutions}).
The corresponding energies are nearly exactly reproduced by the harmonic
approximation, and they are in good agreement with experimental results
(Fig.~\ref{fig:Comparison-to-Experiment}). The harmonic Ansatz greatly
facilitates analysis of interband processes. The last step in our
theory is to analyze the marginal interband scattering within the
harmonic approximation.

SWNTs inherit the degeneracy of the $K$ and $K'$ points from graphene's
dispersion relationship. We refer to the resulting degenerate band
pair (sometimes referred to as valleys) by the band index $n$ for
the nanotube, illustrated in Fig.~\ref{fig:Dispersion}. In semiconducting
tubes the bare gaps are $\Delta_{0}^{\left(n\right)}=2nv_{F}/3R$
($n=1,2,4,5$ for the first four bands) according to the nearest neighbor
tight-binding, zone-folding approximation \cite{Charlier2007,saito1998physical}.
In this relationship $v_{F}$ is the graphene Fermi velocity and $R$
is the tube radius. We assume that the tube radius is not so small
that backscatter and Umklapp processes become relevant \cite{KaneCoulomb1997}.
Reference~\onlinecite{EggerGogolin} derived an effective low-energy
theory for metallic SWNTs using the bosonization technique and we
refer the reader to this work for details on the fermion to boson
mapping. The bosonized reference Hamiltonian, $H_{0}^{\left(n\right)}$,
is the LL Hamiltonian. It includes the free-particle and Coulomb interaction
and is expressed in terms of pairs of dual bosons $\theta_{\nu}^{\left(n\right)}\left(x\right)$
and$\,\phi_{\nu}^{\left(n\right)}\left(x\right)$. For a given band,
$n$, the reference Hamiltonian is a sum of four sectors. 
\begin{equation}
H_{0}^{\left(n\right)}=\frac{1}{2}\sum_{\nu}u_{\nu}\int dx\,\frac{1}{K_{\nu}}\left(\partial_{x}\theta_{\nu}^{\left(n\right)}\right)^{2}+K_{\nu}\left(\partial_{x}\phi_{\nu}^{\left(n\right)}\right)^{2},\label{eq:H0_Bosonic}
\end{equation}
with $\hbar=1$. We follow the conventions of Ref.~\onlinecite{EggerGogolin}
where the $\theta_{\nu}$ fields are associated with density fluctuations,
and the $\phi_{\nu}$ fields act as a corresponding phase. The subscript
$\nu=c\pm,\, s\pm$ indicates charge and spin modes for the sum and
difference from the two degenerate bands. The Coulomb interactions
determine the values of the Luttinger parameters, $K_{\nu}$. Only
the total charge sector, $\nu=c+$, is interacting, with a Luttinger
parameter value less than unity. Typically in nanotubes $K_{c+}\approx0.2$
\cite{EggerGogolin}. The $c+$ sector gives the long wavelength fluctuations
of the total electron density: $\rho\left(x,t\right)=\frac{2}{\sqrt{\pi}}\sum_{n}\partial_{x}\theta_{c+}^{\left(n\right)}\left(x,t\right)$.
The Coulomb interaction increases the velocity for the $c+$ sector,
$u_{c+}\approx v_{F}/K_{c+}$. The remaining sectors have $u_{\nu}=v_{F}$,
and we will refer to these $\nu\ne c+$ sectors as the neutral sectors. 

The resonances in the current-current time correlation function correspond
to peaks in the absorption spectrum. They occur at the gaps, $\tilde{\Delta}^{\left(n\right)}$,
of the $c+$ sector in the fully interacting theory. The gap terms
that are quadratic in fermion fields have a complicated form in the
bosonized representation \cite{Levitov2003}, 
\begin{equation}
H_{\mathrm{gap}}^{\left(n\right)}=\frac{4\Delta_{0}^{\left(n\right)}}{\pi a}\int dx\,\prod_{\nu}\cos\sqrt{\pi}\theta_{\nu}^{\left(n\right)}+\prod_{\nu}\sin\sqrt{\pi}\theta_{\nu}^{\left(n\right)}.\label{eq:BosoHGap}
\end{equation}
The parameter $a$ is the short distance cutoff of the theory \cite{EggerGogolin}.
Under RG each cosine and sine term has a scaling dimension $K_{\nu}/4$
\cite{Gogolin2004}. $H_{\mathrm{gap}}^{\left(n\right)}$ therefore
has the scaling dimension $\sum_{\nu}K_{\nu}/4<1$. A scaling dimension
of 2 would indicate a marginal perturbation, but $H_{\mathrm{gap}}^{\left(n\right)}$,
is far from this value and is relevant. It should not be treated using
perturbation theory. 

This result also implies that the LL line in SWNT is fragile. Metallic
zigzag and chiral tubes can have a curvature induced gap \cite{kane_size_1997}.
These gaps are small, on the order of tens of meVs \cite{Ouyang2001},
but the RG flow moves the system rapidly away from the line of LL
critical points at zero gap. While perturbation theory is unreliable
at $K_{c+}\approx0.2$, the system becomes amenable to a semiclassical
approximation.

$H_{\mathrm{gap}}$ includes only the $\theta_{\nu}^{\left(n\right)}$
fields. One can obtain a description of the gapped tube, $H_{0}^{\left(n\right)}+H_{\mathrm{gap}}^{\left(n\right)}$,
solely in terms of $\theta_{\nu}^{\left(n\right)}$ by going to a
Euclidean action and integrating out the $\phi_{\nu}^{\left(n\right)}$
fields\cite{Giamarchi2004}. The effective action involves all sectors
and describes a coupled sine-Gordon (sG) model. While the action is
highly nonlinear, the sine-Gordon model is one of the most studied
models in quantum field theory \cite{coleman_quantum_1975}(citations
therein). There are an infinite number of discrete ground states corresponding
to the $\theta_{\nu}$ fixed at sine or cosine minima. There are particle-like
excitations, finite in energy and spatial extent, that can be classified
according to their topological charge, $Q_{\nu}=\int_{-\infty}^{\infty}dx\,\partial_{x}\theta_{\nu},$
which is a conserved quantity. In the semiclassical case, the excitations
are solitons that ``traverse'' between adjacent ground states. One
possible solution for the gapped nanotube has for all sectors $\theta_{\nu}\left(x,t\right)=0$
as $x\rightarrow-\infty$ and $\theta_{\nu}\left(x,t\right)=\sqrt{\pi}/2$
as $x\rightarrow+\infty$. This is a composite soliton which traverses
minima between the sine and cosine arguments. Ref.~\onlinecite{Levitov2003}
showed that an electron, as a single particle excitation to be distinguished
from the excitons, can be represented by these composite solitons. 

But the effective action admits other, higher energy, topological
solutions. Consider the case where the neutral sectors are uniform
throughout space with a value $\theta_{\nu}=2m\sqrt{\pi}$ and $\theta_{c+}$
changes from $0$ to $2\sqrt{\pi}$ between the ends of the tube.
This is an example of a pure soliton, three sectors are fixed while
the fourth jumps between cosine minima. Both the composite soliton
and the pure soliton are single-particle excitations. If the soliton
is the analogue of the electron and the antisoliton is the analogue
of the hole, then the bound, topologically neutral soliton-antisoliton
pair is the analogue of the exciton. In sG theory these bound states
are referred to as breathers, illustrated in Fig.~\ref{fig:ClassicalBreatherSolutions}.
Without resorting to numerical simulations\cite{konik2011}, it is
not straightforward to solve for all of the breather solutions in
the effective action. A complete solution to the spectrum of the quantum
field theory would require one to determine all bound states of the
classical field theory corresponding to the quantum action and then
make what is effectively a WKB approximation \cite{Rajaraman1996,Dashen1974a}.
This is beyond the scope of the current work which shows that important,
semiquantitative results can be had without resorting to such measures.

To make progress, we proceed as in Ref.~\onlinecite{konik2011} and
make a mean-field approximation that assumes the $\theta_{\nu}^{\left(n\right)}$
from different sectors are uncorrelated. Within a given band, $n$,
the theory is one of four scalar sG actions, $S^{MF}=S_{0}+S_{\mathrm{gap}}$,
for each boson sector. Written in canonical form \cite{Lukyanov1997},
$S_{\mathrm{gap}}$ for each sector within each band, in the mean-field
approximation, becomes
\begin{equation}
S_{\mathrm{gap}}\left[\theta_{\nu}^{\left(n\right)}\right]=-2\mu_{\nu}^{\left(n\right)}u_{\nu}\int dx\, d\tau\,\cos\sqrt{\pi}\theta_{\nu}^{\left(n\right)}.\label{eq:sG_Coupling}
\end{equation}
 The coupling constants are solved self-consistently, and the $\mu_{c+}^{\left(n\right)}$
then determine the energies of the breathers, $m_{c+}^{\left(n\right)}$
\cite{konik2011,NotesKonikError,Lukyanov1997}. The breather solutions
of the $c+$ sector, formed from pure solitons and antisolitons, are
the semiclassical result that correspond to the excitons of the optical
transitions. 

While the soliton solutions to the $\theta_{c+}$ fields jump between
ground states, the breather solutions oscillate narrowly about a minimum
in the cosine function. In analogy to a particle in a potential well,
$S_{gap}$ can be replaced by its harmonic approximation, $1-\cos\sqrt{\pi}\theta_{c+}\sim\theta_{c+}^{2}$.
In this form, the decoupled sG action describes a theory of free massive
bosons, or quantum harmonic oscillators, whose dispersion relationship
is determined by the full gaps, \cite{Giamarchi2004} 
\begin{equation}
\tilde{\Delta}^{\left(n\right)}=u_{c+}\left[2\pi\mu_{c+}^{\left(n\right)}K_{c+}\right]^{1/\left(2-K_{c+}/4\right)}.\label{eq:HarmonicMassTerm}
\end{equation}
The application of the RG in nanoscale systems often results in cutoff-dependent
parameters \cite{KaneElectron2004}. The same is true here. The coupling
constants $\mu_{c+}^{\left(n\right)}$ are inherently scale dependent.
Their values depend on the short distance cutoff $a$ {[}see Eq.~(\ref{eq:BosoHGap}){]}.
We use a normalization scheme that first determines the value for
$\tilde{\Delta}^{\left(n\right)}$ in the noninteracting case, $K_{c+}=1$.
The ratio $\tilde{\Delta}^{\left(n\right)}/\Delta_{0}^{\left(n\right)}$
gives a single normalization factor applied to solutions with $K_{c+}=0.2$.
The full gaps $\tilde{\Delta}^{\left(n\right)}$ in the quantum harmonic
approximation are within half a percent of the breather energies $m_{c+}$
evaluated in the semiclassical approximation. The correlation length
in the free massive boson theory is inversely proportional to the
mass term, $\tilde{\Delta}^{\left(n\right)}$, and we find that the
ratio of correlation lengths follows from the ``ratio problem,''
$\xi^{\left(1\right)}/\xi^{\left(2\right)}\approx1.78$ \cite{KaneRatio2003}.
When estimated from the semiclassical particle density (Fig.~\ref{fig:ClassicalBreatherSolutions})
the ratio of the full widths at half maximum, gives $\xi^{\left(1\right)}/\xi^{\left(2\right)}\approx1.76$,
which implies that the {\em dynamics} for the lowest energy excitations
in the quantum harmonic approximation are nearly indistinguishable
from those evaluated with semiclassical methods.

The full gaps from Eq.~(\ref{eq:HarmonicMassTerm}) give the optical
transition energies. Just as in Ref. \cite{KaneElectron2004}, we
plot them alongside the experimental data in Fig.~\ref{fig:Comparison-to-Experiment}.
Since we set $K_{c+}=0.2$ for the interacting case and we fix the
short distance cutoff to be the carbon-carbon distance, our theory
has a single free parameter: the radius $R$, or equivalently, the
bare gaps $\Delta_{0}^{\left(n\right)}$. The agreement between our
theory and experiment is quantiative for the $E_{11}$ transition
in large radius tubes and semiquantiative otherwise. The largest disagreements
between our theory and experiment occur when $R$ is small, which
is to be expected for a field-theoretical treatment. For small radius
tubes, short wavelength behavior and a host of marginal and irrelevant
terms, ignored in the field theory, contribute. 

In addition to our calculated optical transition energies in Fig.~\ref{fig:Comparison-to-Experiment}
we also show theoretical results from Ref.~\onlinecite{KaneElectron2004}.
The tight-binding prediction is that $\Delta_{0}\propto R^{-1}$.
Kane and Mele \cite{KaneElectron2004} used RG arguments and Bethe-Salpeter
calculations and found a logarithmic dependence on the tube radius
$\tilde{\Delta}/\Delta_{0}\propto\ln R$. The results from the harmonic
approximation show a power law relationship $\tilde{\Delta}/\Delta_{0}\propto R^{\gamma}$
with a small nonuniversal exponent $\gamma=\left(1-K_{c+}\right)/\left(5-K_{c+}\right)$.

Having identified and analyzed the strongly interacting parts of the
theory, we turn to the interband scattering dynamics. In the Fourier
basis, the vertex of the Coulomb interaction on a cylinder of radius
$R$ and wave vector $q$ behaves as $V_{0}\left(q\right)\sim\left|\ln(qR)\right|$
for intraband scattering, and $V_{I}\left(q\right)\sim1/2+q^{2}R^{2}$
for scattering between adjacent bands at small $q$. A $3\,\mu\mathrm{m}$
length, $1\,\mathrm{nm}$ diameter tube has $V_{0}/V_{I}\sim20$.
For simplicity we look at scattering between the lowest two energy
bands. The interband term involves only densities from the total charge
sector in each band and contains no anharmoncities, 
\begin{eqnarray}
S_{I}\left[\theta_{c+}\right] & = & \frac{1}{2}\frac{4V_{I}}{\pi}\int dx\,\partial_{x}\theta_{c+}^{\left(1\right)}\left(x\right)\,\partial_{x}\theta_{c+}^{\left(2\right)},\label{eq:H_interacting}
\end{eqnarray}
and is exactly solvable in the harmonic approximation. It leads to
a small hybridization between bands, equivalent to a redefinition
of the normal modes, but nothing more. The strongly-interacting low-dimensional
aspect of the Coulomb interaction is manifest between carriers within
the same band, but between carriers in different bands, it is a relatively
small effect. This conclusion is consistent with other atomistic calculations
based on the Bethe-Salpeter equation \cite{PerebeinosScaling2004}.

The boson fields can be quantized in terms of boson creation and annihilation
operators \cite{Giamarchi2004,Voit1995}:
\begin{equation}
\partial_{x}\theta_{\nu}^{\left(n\right)}\left(x\right)\sim\sum_{k\ne0}\, e^{-a\left|k\right|/2}e^{-ikx}\left(b_{k}^{\left(n\right)\dagger}+b_{-k}^{\left(n\right)}\right).\label{eq:ThetaModeExpansion}
\end{equation}
The interband scattering includes $b_{k}^{\left(1\right)\dagger}b_{k}^{\left(2\right)}$,
and its Hermitian conjugate, but no higher order terms that could
lead to multiple exciton generation. Reference~\onlinecite{Balents2000a}
also considers an interband forward scattering model in carbon nanotubes,
they find a MEG-like term in the Hamiltonian but it is determined
to be irrelevant. It is significant that semiconducting SWNTs are
well described as a system of harmonic oscillators (Fig.~\ref{fig:Comparison-to-Experiment}).
Within our approximations, long wavelength interband scattering does
not lead to multiple exciton generation. 

In this work we show that the optical transition energies for semiconducting
SWNTs can be determined with a minimal number of parameters using
a forward scattering bosonization model, mean-field theory, and a
harmonic approximation. The multiband SWNT reduces to a system of
weakly coupled harmonic oscillators. We have shown that the LL phase
of  metallic SWNTs are unstable in the presence of a bare gap of any
size. These gaps may be introduced by curvature effects \cite{Saito2000},
magnetic fields \cite{Levitov2003}, or strain \cite{Cronin2005}.
Under the influence of a bare gap, correlation lengths become finite.
The electron fluctuations, localized in space, may be thought of as
quasiparticles. This may be related at a fundamental level to the
success of quasiparticle based Bethe-Salpeter theories for excitons
in SWNTs. 

This work was partially supported through a grant from CRSP (Award
KXRE-9-99004-09). JDE would like to thank the University of Colorado
for generous startup funds.


\begin{thebibliography}{40}
\bibitem [{Not()}]{NotesKonikError}%
  \BibitemOpen
  \href@noop {} {}\bibinfo {note} {{O}ur derivation result for the breather
  energy differs from that in Reference \onlinecite{konik2011}. In their
  Equation (6) the last gamma function should have argument
  $\frac{1}{2}+\frac{\xi_{c+}}{2}$ (not $\frac{1}{2}+\xi$), and the definition
  of the scaling dimension should be $\alpha\equiv K_{c+}/4$.}\BibitemShut
  {Stop}%
\end{thebibliography}
\end{document}